# Directional dependence of the plasmonic gain and nonreciprocity in drift-current biased graphene

Tiago A. Morgado[1], Mário G. Silveirinha[2,*]

[1]Instituto de Telecomunicações and Department of Electrical Engineering, University of Coimbra, 3030-290 Coimbra, Portugal

[2]University of Lisbon–Instituto Superior Técnico and Instituto de Telecomunicações, Avenida Rovisco Pais, 1, 1049-001 Lisboa, Portugal

*E-mail:* tiago.morgado@co.it.pt, mario.silveirinha@co.it.pt

**Abstract**

Here, we investigate the nonreciprocal propagation and amplification of surface plasmons in drift-current biased graphene, using both Galilean and relativistic-type Doppler shift transformations of the graphene's conductivity. Consistent with previous studies, both conductivity models predict strongly nonreciprocal propagation of surface plasmons due to the drag effect caused by the drifting electrons. In particular, the Galilean Doppler shift model leads to stronger spectral asymmetries in the plasmon dispersion with regimes of unidirectional propagation. Remarkably, it is shown that both conductivity models predict regimes of nonreciprocal plasmon amplification in a wide angular sector of in-plane directions when the drift-current biased graphene sheet is coupled to a plasmonic substrate (namely, SiC), with the plasmon amplification rate being substantially higher for the relativistic Doppler shift model.

[*] To whom correspondence should be addressed: E-mail: mario.silveirinha@co.it.pt

- **Introduction**

Reciprocity is an intrinsic property of usual passive, linear, and time-invariant photonic systems [1-5]. The Lorentz reciprocity principle enforces that the response of conventional photonic platforms stays the same when the position of the source and detector are interchanged [2-4]. This has important practical implications, as it implies that conventional optical systems are bi-directional.

Breaking reciprocity is of uttermost importance for photonic technologies, as it enables one-way light propagation and optical isolation. The most common way to do so is by biasing magneto-optical materials (ferrites or other iron garnets) with a static magnetic field [6-8], which gives rise to gyrotropic nonreciprocal responses. However, the need of an external bulky biasing circuit, as well as the relatively weak gyrotropic responses in the terahertz and optical regimes, hinder the usefulness of these nonreciprocal "magnetic" solutions in highly-integrated photonic systems. Due to this, great efforts have been devoted to developing magnetic-free nonreciprocal photonic solutions that can be directly incorporated in nanophotonic systems [9]. Such alternative nonreciprocal solutions include time-modulated systems [10-15], active electronic systems [16-17], optomechanical cavities [18-19], moving media [20-23], non-Hermitian platforms [24-26], systems that exploit dynamic nonlinear effects [27-32], among others.

An alternative and interesting route to break the reciprocity at the nanoscale is through the biasing of graphene with a drift electric current [33-38]. Taking advantage of the ultrahigh electron mobility of graphene, we have theoretically and numerically demonstrated in [35,39] that a drift-current biased graphene sheet may imitate the optical response of a moving medium [22,40], also leading to a breaking of the time-reversal symmetry and strong nonreciprocal responses. Remarkably, the nonreciprocal response of drift-current biased graphene was experimentally demonstrated recently [41-42]. Furthermore, we have theoretically and numerically demonstrated that systems formed by a drift-current biased graphene sheet coupled to a plasmonic-type material slab (e.g., a SiC substrate) may enable active (non-Hermitian) responses wherein the graphene plasmons are pumped by the drifting electrons [39, 43-44], and which can lead to regimes of loss compensation, plasmonic amplification and spasing [44].

The objective of this work is to study in detail how the nonreciprocal and non-Hermitian plasmonic effects in the drift-current biased graphene depend on the direction of propagation of the SPPs. In previous studies [35, 44], we considered two-dimensional (2D) geometries wherein the graphene sheet is characterized by a scalar surface conductivity $\sigma_g(\omega, \mathbf{k})$. Here we extend the analysis to fully three-dimensional (3D) scenarios wherein the electromagnetic response of graphene is described by a surface conductivity tensor $\bar{\bar{\sigma}}_g(\omega, \mathbf{k})$. In particular, we analyze in detail the propagation of SPPs in systems formed by a 2D drift-current biased graphene sheet deposited on dielectric (SiO$_2$ or h-BN) or plasmonic (SiC) substrates when excited by near-field electric dipolar sources with different polarizations. We use both the Galilean [39] and the relativistic [33-34, 36] Doppler shift models to characterize the conductivity of the drift-current biased graphene, and find that for the case of the SiC substrate there is a wide angular sector within which the SPPs may be amplified due to the gain in the material.

It should be noted that the drift-induced plasmon drag was recently experimentally verified in [41-42]. The experiments were performed in the kinetic or collisionless regime, $\omega \gg \gamma_{e-e}$, with $\gamma_{e-e}$ the electron-electron (e-e) collision frequency. It was verified that in such a regime the relativistic Doppler-shift model fits very well the experimental data. On the other hand, as discussed in detail in Ref. [43], the Galilean model may be applicable in the regime where the electron-electron collisions dominate ($\omega \ll \gamma_{e-e}$), leading to a shifted Fermi distribution that describes a fluid that moves as a whole with constant speed [45], analogous to a moving medium. Here, we find that while the Galilean model generically predicts a stronger spectral asymmetry in the plasmons dispersion, it is the relativistic model that predicts the strongest gain.

- **Results and Discussion**

Figure 1 illustrates the nanostructure under study. It consists of a graphene sheet deposited on the top of a substrate and traversed by a drift electric current created by a static voltage drop applied across the sheet. We assume that the region above graphene is air.

In the absence of an electron drift, the graphene sheet can be characterized by the surface conductivity tensor

$$\overline{\boldsymbol{\sigma}}_{\mathrm{g}} = \sigma_{\mathrm{L}} \left( \hat{\mathbf{k}}_{\mathrm{t}} \otimes \hat{\mathbf{k}}_{\mathrm{t}} \right) + \sigma_{\mathrm{T}} \left( \hat{\mathbf{z}} \times \hat{\mathbf{k}}_{\mathrm{t}} \right) \otimes \left( \hat{\mathbf{z}} \times \hat{\mathbf{k}}_{\mathrm{t}} \right), \tag{1}$$

where $\mathbf{u} \otimes \mathbf{v}$ represents the dyadic product (tensor product) of two generic vectors $\mathbf{u}$ and $\mathbf{v}$, $\hat{\mathbf{k}}_{\mathrm{t}} = \mathbf{k}_{\mathrm{t}}/|\mathbf{k}_{\mathrm{t}}|$ (with $\mathbf{k}_{\mathrm{t}} = k_x \hat{\mathbf{x}} + k_y \hat{\mathbf{y}}$ being the transverse wave vector and $|\mathbf{k}_{\mathrm{t}}| \equiv k = \sqrt{k_x^2 + k_y^2}$), and $\sigma_{\mathrm{L}}(\omega, k)$ and $\sigma_{\mathrm{T}}(\omega, k)$ are the nonlocal surface conductivities of the graphene sheet for longitudinal (with in-plane electric field parallel to the wave vector $\mathbf{k}$) and transverse (with in-plane electric field perpendicular to the wave vector $\mathbf{k}$) excitations, respectively. The intraband longitudinal and transverse conductivities are given in the Supplemental Material [46].

The Galilean Doppler shift model assumes that the electrons move collectively as a whole with a constant velocity [39,43]. In such a case, the electromagnetic response of a graphene sheet biased with a drift current is effectively equivalent to that of a graphene sheet without drifting electrons in translational nonrelativistic motion. In particular, analogous to a moving medium [40], for this model the reflection matrix of the graphene biased with a drift current ($\tilde{\mathbf{R}}$) is related to the reflection matrix of the bare graphene without drifting electrons ($\mathbf{R}$) as $\tilde{\mathbf{R}}(\omega, k_x, k_y) \approx \mathbf{R}(\tilde{\omega}, k_x, k_y)$, where $\tilde{\omega} = \omega - k_x v_0$ is the Doppler-shifted frequency and $v_0$ is the drift velocity of the electrons on the graphene sheet along the *x*-direction. Thus, the response of the material without the drift current fully determines the electromagnetic response with the drift current. Notice that $\tilde{\mathbf{R}}$ takes into account the spatial dispersion inherent to the graphene's response and the difference between the longitudinal and transverse conductivities. The reflection matrix for a graphene sheet without drifting electrons is derived in the Supplemental Material [46].

Alternatively, the response of the drift-current biased graphene can be modeled by a relativistic Doppler shift model [33-34, 36], which as previously mentioned was shown to fit well the experimental data in the kinetic regime [41-42]. For simplicity, in this case, we assume that the graphene sheet with no drift is characterized by a scalar conductivity $\sigma_g(\omega, k) = \sigma_L(\omega, k)$ rather than by the conductivity tensor (1) as in the Galilean model, so that the graphene conductivity in the presence of a drift current is given by $\sigma_g^{\text{Rel}}(\omega, k) \approx (\omega / \tilde{\omega}) \sigma_g(\tilde{\omega}, \tilde{k})$, with $\tilde{\omega} = \gamma(\omega - k_x v_0)$, $\tilde{k}_x = \gamma(k_x - \omega v_0 / v_F^2)$, and $\gamma = 1/\sqrt{1 - v_0^2 / v_F^2}$ the graphene Lorentz factor.

To begin with, we investigate the dispersion and damping properties of the surface plasmons polaritons (SPPs) supported by a graphene sheet biased with a drift-current and deposited on the top of a dielectric substrate. Figure 2(a) depicts the dispersion of the graphene SPPs copropagating ($k_x' > 0$; the single prime denotes the real part of the wave number) and counter-propagating ($k_x' < 0$) with the drifting electrons, calculated using the Galilean (solid curves) and the relativistic (dashed lines) models. Both models predict that the drift-current biasing of graphene causes a symmetry breaking in the SPP dispersion, being the spectral asymmetry clearly stronger for the Galilean model. In addition, Fig. 2(b) shows the ratio between the attenuation constants of the SPPs counter-propagating ($k''_{x,-}$; the two primes denote the imaginary part of the wave number) and copropagating ($k''_{x,+}$) with the drifting electrons. Clearly, the counter-propagating SPPs ($-x$ direction) are more attenuated than the copropagating plasmons ($+x$ direction). The discrepancy between the attenuation constants of the counterpropagating and copropagating plasmons is more evident in the Galilean model.

To characterize the directional properties of the SPPs, we calculated the isofrequency contours for different drift velocities $v_0$. In the absence of a drift-current ($v_0 = 0$), the isofrequency contour is circular and centered at the origin (($k'_x, k'_y$) = (0,0)) [see green curve in Fig. 2(c); here, $k_l = k'_l + ik''_l$, with $l = x, y$]. This means that for every positive $k$-solution there is also an equivalent $-k$ solution, in agreement with the reciprocity of the system. Moreover, the attenuation constant $\alpha$ is the same for all directions of propagation in the 2D graphene surface.

With a drift current biasing, the Galilean model predicts an isofrequency contour that is elliptical and not centered at the origin [see blue curve in Fig. 2(c)]. In addition, the attenuation constant is strongly dependent on the direction of propagation. In contrast, for the relativistic model the isofrequency contours are circular, but with the contour center displaced from the origin. The circular shape of the isofrequency contours is explained by the fact that the graphene conductivity is a scalar in the relativistic model. Moreover, the attenuation constant is also $k$-dependent [see purple curve in Fig. 2(c)]. Besides the different isofrequency contours, the two models predict rather different SPP dispersions, especially for the SPPs in the left **k**-semiplane (with $k'_x < 0$ ). The Galilean model predicts SPPs in the left **k**-semiplane with much larger wavenumber (i.e., the SPPs are more confined) than the relativistic model, in full agreement with the results of Fig. 2(a).

The asymmetry of the isofrequency contours relative to the origin and the direction dependent attenuation are fingerprints of the reciprocity breaking in the system. In particular, consistent with the results of Figs. 2(a)-(b), it is found that in the presence of a drift-current biasing, the SPPs in the right **k**-semiplane (i.e., with $k'_x > 0$) are significantly less confined and attenuated than the plasmons in the left **k**-semiplane (with $k'_x < 0$) [35, 37-38].

Next, we investigate the excitation of the SPPs by a near-field dipole source. The emitted fields are calculated using a theoretical formalism based on Sommerfeld-type integrals. The details can be found in the Supplemental Material [46]. We start by considering configurations where the drift-current biased graphene sheet is deposited on the top of a dielectric substrate with relative permittivity $\varepsilon_{r,s} = 4$ (SiO$_2$ or h-BN). Figures 3(a)(*i*)-(*iii*) show time snapshots of the out-of-plane component of the electric field ($E_z$) on the surface of the graphene sheet excited by a linearly polarized vertical dipole with electric dipole moment $\mathbf{p}_e = p_{e,0} \hat{\mathbf{z}}$ and placed 30 nm above the graphene sheet. Consistent with the dispersion diagram and the isofrequency contour of Fig. 2, without a drift-current bias (Fig. 3(a)(*i*)), there is no preferred direction of propagation for the graphene SPPs excited by the vertical dipole. The SPP wavefronts are circular and the field amplitude is constant on each wavefront. Quite differently, with a drift-current biasing (Fig. 3(a)(*ii*)-(*iii*)), the SPP propagation along the direction of the drifting electrons (+*x* direction) is clearly favored. In agreement with Refs. [35, 37-38] and Fig. 2, the Galilean model (Fig. 3(a)(*ii*)) predicts a unidirectional SPP propagation regime along the *x*-axis. In fact, the SPPs co-propagating with the drifting electrons (along the $+x$ direction) are much less attenuated than the counter-propagating SPPs (along the $-x$ direction). Furthermore, the Galilean model predicts that the counter-propagating SPPs have extremely large wave numbers, which makes their excitation more difficult.

On the other hand, for the relativistic Doppler shift model the SPP propagation is not strictly unidirectional [see Fig. 3(a)(*iii*)]. Indeed, the asymmetry of the propagation and attenuation constants dispersion is evidently less pronounced for the relativistic model [38] than for the Galilean model, as shown in Fig. 2.

We also investigated configurations in which the drift-current biased graphene is excited by an electric dipole with circular polarization ($\mathbf{p}_e = p_{e,0}(\hat{\mathbf{y}} + i\hat{\mathbf{z}})$). In this case, due to the spin-momentum locking [47], the SPPs are asymmetrically excited in the unbiased graphene sheet [48]. For $\mathbf{p}_e = p_{e,0}(\hat{\mathbf{y}} + i\hat{\mathbf{z}})$ the SPPs are mostly launched towards the $y > 0$ semi-plane (Fig. 3(b)(*i*)). Both theoretical models predict that with a drift current bias, the graphene SPPs are dragged by the drifting electrons towards the $+x$-direction [see Figs. 3(b)(*ii*)) and 3(b)(*iii*) for the Galilean and relativistic models, respectively].

Let us now consider that the positive permittivity substrate is replaced by a plasmonic-type (negative permittivity) substrate, such as silicon carbide (SiC) ($\varepsilon_{r,s} = \varepsilon_{r,SiC}(\omega)$). SiC is characterized by the dielectric function reported in [49-50]. Here we concentrate on the frequency range $22.78 < f$ [THz] $< 27.82$ where the SiC has a plasmonic response characterized by $\text{Re}\{\varepsilon_{SiC}\} < 0$. The graphene-SiC system supports deeply confined SPPs, which result from the hybridization of the graphene and SiC plasmons [44]. Due to the drift current, the plasmonic response can be strongly non-Hermitian as the SPPs may be pumped by the kinetic energy of the drifting electrons [43-44].

Figure 4(a)-(b) depicts the wavelength and the attenuation constant of the graphene-SiC plasmons copropagating with the drifting electrons as a function of the frequency for different drift velocities $v_0$, using both the Galilean (solid curves) and the relativistic (dashed curves) Doppler shift models. It is seen from Fig. 4(a) that the SPP wavelength decreases as the frequency increases (i.e., the SPPs become more confined), as expected. In addition, Fig. 4(a) shows that the Galilean model predicts SPPs with larger wavelength than the relativistic model, especially for lower frequencies. Notably, Fig. 4(b) shows that for sufficiently large drift velocities, the attenuation constant of the SPPs co-propagating with the drifting electrons ($+x$ direction) may vanish ($\alpha = 0$) or even become negative ($\alpha < 0$). In the latter situation the graphene system behaves as a distributed amplifier. The relativistic model predicts weaker attenuation and larger amplification rates [see dashed curves in Fig. 4(b)].

Figure 4(c) depicts the isofrequency contours of the SPPs supported by a drift-current biased graphene sheet (with $v_0 = 0.6 v_F$) deposited on the top of a SiC substrate. Notably, both models predict a negative attenuation constant for the SPPs inside an angular sector in the right semiplane ($k'_x > 0$) centered about the $k'_x$-axis, so that the SPPs are amplified in this angular sector. Interestingly, the relativistic model predicts an amplification sector with a larger angular width (around $89°$) than the Galilean model (about $63°$), as well as a larger amplification rate. On the other hand, both models predict that the SPPs in the left semiplane ($k'_x < 0$) are strongly attenuated. It is curious that for counter-propagating plasmons the Galilean model yields the largest attenuation, whereas for co-propagating plasmons the relativistic model yields the largest gain.

Figures 5(a-c) present time snapshots of the *x*-component of the electric field on the surface of a drift-current biased graphene sheet (with $\mu_c = 0.1$ eV) deposited on a SiC substrate and excited by a linearly polarized vertical dipole ($\mathbf{p}_e = p_{e,0}\hat{\mathbf{z}}$). Figure 5(a) shows the results calculated using the Galilean model for a drift velocity $v_0 = 0.6v_F$. Remarkably, the SPPs co-propagating with the drifting electrons are amplified, whereas the counter-propagating SPPs are so strongly attenuated that the SPP propagation is effectively unidirectional. Such a regime of unidirectional SPP propagation and amplification is in fully agreement with the results of Fig. 4. Furthermore, it is shown in the Supplementary Material [46] that similar drift-induced unidirectional propagation and amplification regimes can be attained with dipoles with different polarizations. The gain (non-Hermitian) response emerges due to a negative Landau damping effect that enables the transfer of kinetic energy from the drifting electrons to the highly confined graphene-SiC plasmons [39, 43-44].

The results obtained with the relativistic Doppler shift theory are shown in Fig. 5(b-c). Clearly, for the same drift velocity $v_0 = 0.6v_F$, the relativistic model predicts a significantly stronger SPP amplification in a wider angular range than the Galilean model [see Figs. 5(a-b)], consistent with Fig. 4. Interestingly, for the same gain per unit of propagation length, the drift velocity in the relativistic model can be about 14% smaller than in the Galilean model [see Figs. 5(c-d)].

Next, we show how by tuning the chemical potential it is possible to control the plasmon amplification rate. Figure 6 shows the attenuation constant of the graphene-SiC SPPs copropagating with the drifting electrons as a function of the frequency and of the chemical potential $\mu_c$, for two different drift velocities $v_0$. The results were calculated using the Galilean model. We focus on the frequency range wherein $\text{Re}\{\varepsilon_{\text{SiC}}\} < 0$.

Figure 6 demonstrates that by increasing the chemical potential, it is possible to boost the gain and achieve plasmon amplification for lower frequencies and smaller drift velocities. In principle, the drift velocities considered in Fig. 6 (namely, $v_0 = 0.2 v_F$) are within reach of an experimental realization [41].

- **Conclusions**

In this work, we have theoretically studied the directional properties of the nonreciprocal and non-Hermitian plasmonic effects in fully 3D drift-current biased graphene sheets deposited on a dielectric or plasmonic (SiC) substrate. The effect of the drift current bias on the surface conductivity of graphene is modeled using either the Galilean Doppler shift model [39] or the relativistic Doppler shift model [33-34, 36]. In the case of the Galilean model, we take into account that the graphene electrons interact differently with transverse and longitudinal waves. Both models predict highly asymmetric plasmon dispersions with the SPPs propagating almost solely along the direction of the drift velocity. The asymmetry is more pronounced for the Galilean model and leads to unidirectional SPP propagation. In addition to the strong nonreciprocity, both models predict a qualitatively similar direction-dependent plasmonic amplification when the graphene sheet is placed on top of a SiC substrate. Interestingly, the relativistic Doppler shift model (which was shown to agree with experimental data in the kinetic regime) leads to larger amplification rates than the Galilean model, and to a wider angular range of directions wherein the intrinsic material attenuation is supplanted by the material gain. Therefore, our results suggest that drift-current biased graphene is a promising solution to engineer non-Hermitian active platforms in the terahertz and infrared ranges.

**Acknowledgments:** This work was partially funded by the IET under the A F Harvey Prize, by the Simons Foundation under the award 733700 (Simons Collaboration in Mathematics and Physics, "Harnessing Universal Symmetry Concepts for Extreme Wave Phenomena"), and by Instituto de Telecomunicações (IT) under project UIDB/50008/2020. T. A. Morgado acknowledges FCT for research financial support with reference CEECIND/04530/2017 under the CEEC Individual 2017, and IT-Coimbra for the contract as an assistant researcher with reference CT/No. 004/2019-F00069.

**Figures**

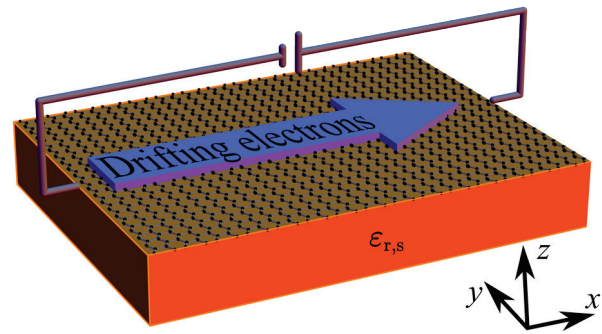

**Figure 1. Drift-current biased graphene sheet.** A graphene sheet deposited on the top of a substrate (with relative permittivity $\varepsilon_{r,s}$) is biased with a drift-electric current. The region above graphene is air.

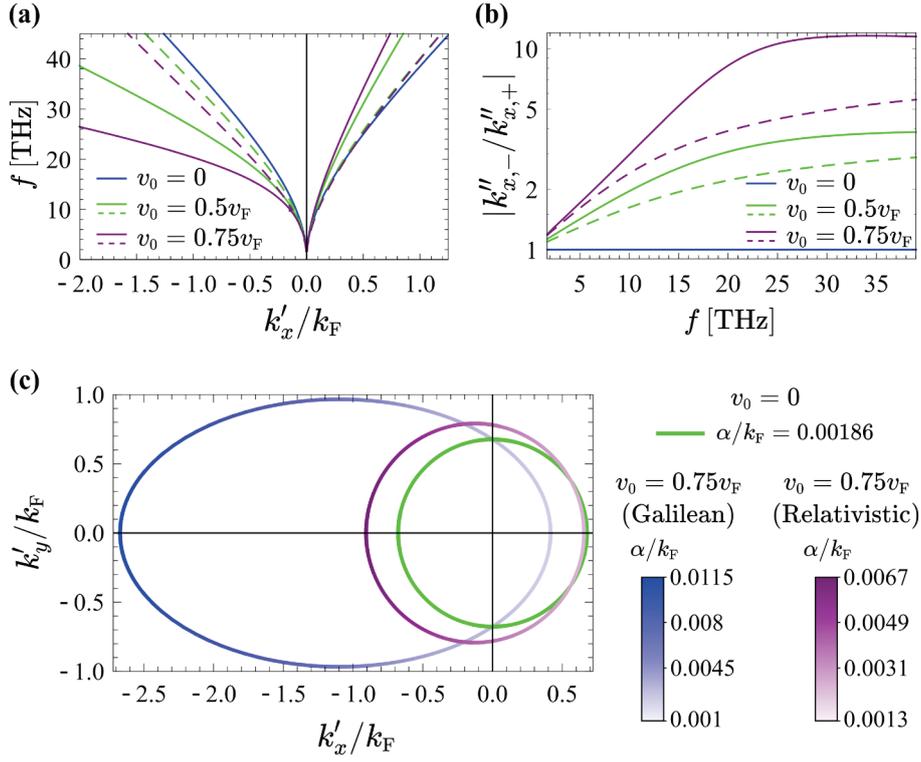

**Figure 2. SPPs dispersion and isofrequency contours for a nondispersive dielectric substrate.** (a) Dispersion of the SPPs, and (b) ratio between the attenuation constants (in logarithmic scale) for propagation along the $-x$ and $+x$ directions as a function of the frequency, for several drift velocities $v_0$; solid curves: Galilean model; dashed curves: relativistic model. (c) Isofrequency contours of the SPPs for $\omega/(2\pi) = 30$ THz; the green curve is calculated without a drift current bias ($v_0 = 0$), and the blue and purple curves with a drift current bias ($v_0 = 0.75 v_F$), using the Galilean model (blue curve) and the relativistic model (purple curve). The color gradient represents the SPP attenuation constant ($\alpha = k'' \, \mathrm{sgn}(k')$). In all the results $\mu_c = 0.1$ eV, $\tau = 1.7$ ps, and $\varepsilon_{r,s} = 4$.

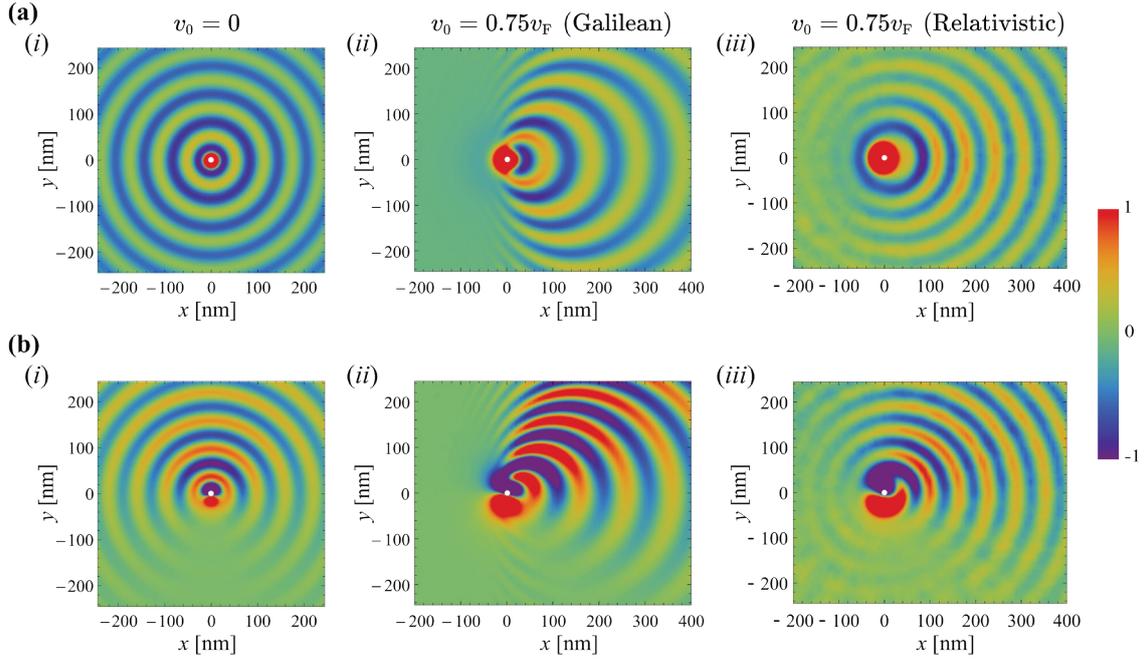

**Figure 3. SPP excitation by an electric dipole.** The graphene sheet stands on the top of a dielectric substrate with $\varepsilon_{r,s}=4$. (a-b) Time snapshots of the z-component of the electric field $E_z$ (in arbitrary unities) calculated on the surface of the graphene sheet for an electric dipole standing at the point $(x,y,z)=(0,0,30)$ nm. (a) Vertical dipole with $\mathbf{p}_e = p_{e,0}\hat{\mathbf{z}}$ (C m); (b) Circularly polarized dipole with $\mathbf{p}_e = p_{e,0}(\hat{\mathbf{y}}+i\hat{\mathbf{z}})$ (C m). (*i*) Without a drift current bias ($v_0=0$); (*ii-iii*) with a drift current bias ($v_0=0.75v_F$), (*ii*) obtained using the Galilean Doppler shift model and (*iii*) the relativistic Doppler shift model. In all the panels the frequency of operation is $\omega/(2\pi)=30$ THz, $\mu_c=0.1$ eV and $\tau=1.7$ ps.

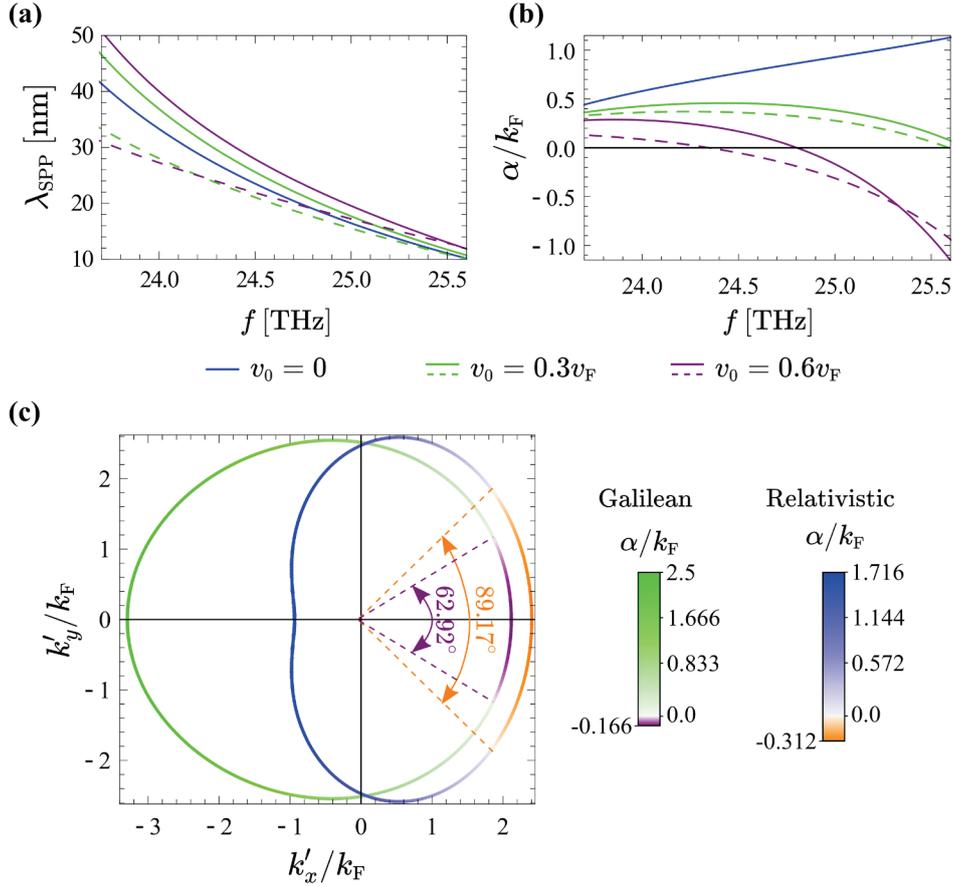

**Figure 4. SPP wavelength, attenuation constant, and isofrequency contours, when the graphene sheet stands on the top of a plasmonic substrate.** (a) Guided wavelength and (b) attenuation constant ($\alpha = k'' \operatorname{sgn}(k')$) for the SPPs copropagating with the drifting electrons as a function of the frequency, for several drift velocities $v_0$. (c) Isofrequency contours of the SPPs for $\omega/(2\pi) = 25$ THz and $v_0 = 0.6 v_F$; the green curve is calculated using the Galilean model, whereas the blue curve is calculated using the relativistic model. The color gradient represents the SPP attenuation constant $\alpha$. The angular sectors wherein the plasmons experience gain (i.e., $\alpha < 0$) are delimited by the dashed lines. In all the panels $\mu_c = 0.1$ eV, $\tau = 1.7$ ps, and $\varepsilon_{r,s} = \varepsilon_{r,\text{SiC}}$.

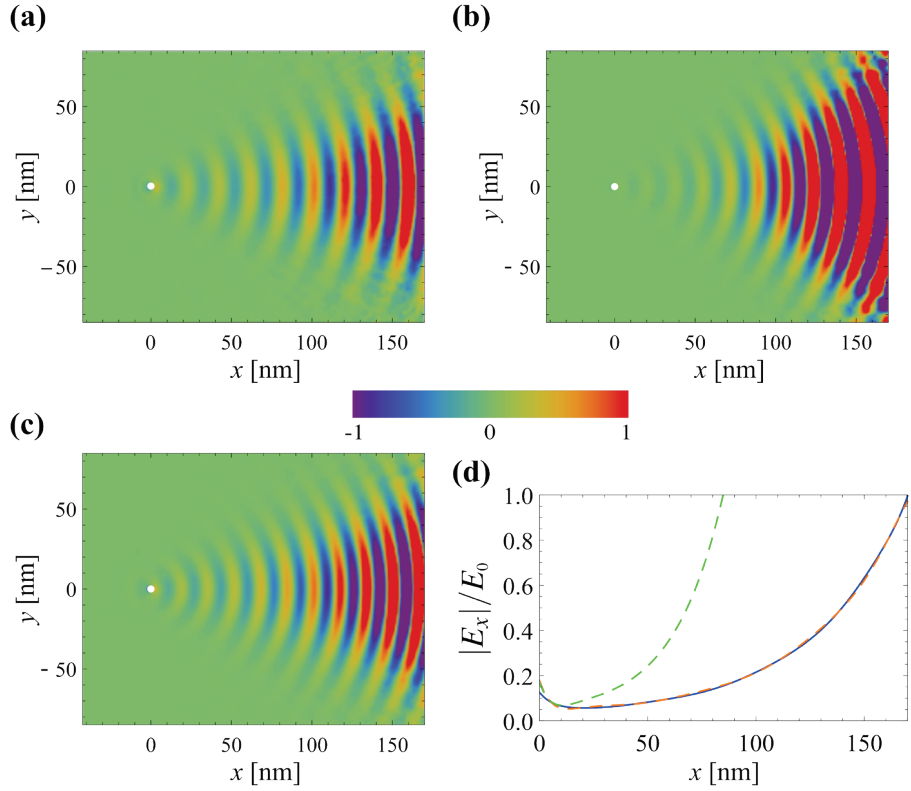

**Figure 5. SPP excitation by an electric dipole with the graphene sheet deposited on a SiC substrate.** (a-c) Time snapshots of the $x$-component of the electric field $E_x$ (in arbitrary unities) on the surface of the graphene sheet for an electric dipole with $\mathbf{p}_e = p_{e,0}\hat{\mathbf{z}}$ (C m) placed 1 nm above the interface. (a) Results calculated using the Galilean model with $v_0 = 0.6 v_F$; (b-c) results calculated using the relativistic model with (b) $v_0 = 0.6 v_F$ and (c) $v_0 = 0.516 v_F$. (d) Amplitude of the $x$-component of the electric field $E_x$ as a function of $x$ and for $(y,z)=(0,0)$, obtained using the Galilean model (blue solid line) and the relativistic model (dashed lines), for $v_0 = 0.6 v_F$ (blue solid and green dashed lines) and $v_0 = 0.516 v_F$ (orange dashed line). In all the panels the frequency of operation is $\omega/(2\pi) = 25$ THz, $\mu_c = 0.1$ eV, $\tau = 1.7$ ps, and $\varepsilon_{r,s} = \varepsilon_{r,\text{SiC}}$.

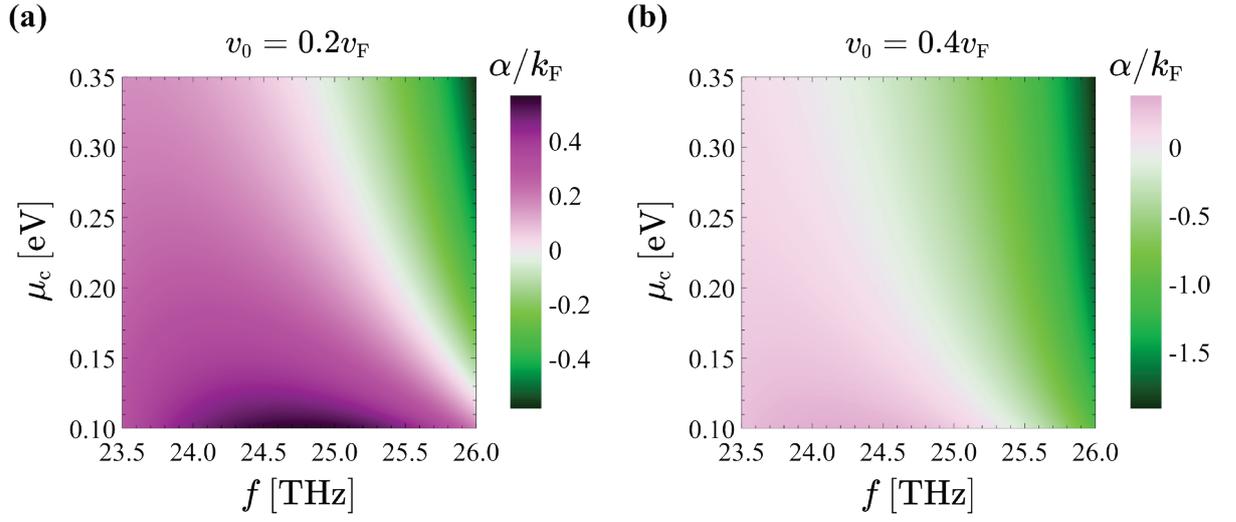

**Figure 6. Parametric study of the SPP attenuation constant.** Attenuation constant ($\alpha = k'' \operatorname{sgn}(k')$) for the SPPs copropagating with the drifting electrons as a function of the frequency and chemical potential $\mu_c$. In all the panels $\tau = 1.7$ ps and $\varepsilon_{r,s} = \varepsilon_{r,\text{SiC}}$.

# Supplemental Material for the Manuscript

# "Directional dependence of the plasmonic gain and nonreciprocity in drift-current biased graphene"

Tiago A. Morgado[1], Mário G. Silveirinha[2*]

[1]*Instituto de Telecomunicações and Department of Electrical Engineering, University of Coimbra, 3030-290 Coimbra, Portugal*

[2]*University of Lisbon–Instituto Superior Técnico and Instituto de Telecomunicações, Avenida Rovisco Pais, 1, 1049-001 Lisboa, Portugal*

*E-mail:* tiago.morgado@co.it.pt, mario.silveirinha@co.it.pt

In the supplementary note *A*) we present the surface conductivity tensor that characterizes the nonlocal intraband response of the bare graphene sheet. In the supplementary note *B*), we derive the reflection and transmission matrices for a two-dimensional (2D) material. The supplementary note *C*), derives the electromagnetic fields radiated by an electric dipole source nearby the 2D material. Finally, in supplementary note *D*), we present a more complete study of the graphene SPPs excited by emitters with different polarizations for the SiC substrate.

## A. Surface conductivity tensor of a graphene sheet

The surface conductivity that characterizes an unbiased (i.e., without drifting electrons) graphene sheet can be written as:

$$\overline{\overline{\sigma}}_g = \sigma_L \left( \hat{\mathbf{k}}_t \otimes \hat{\mathbf{k}}_t \right) + \sigma_T \left( \hat{\mathbf{z}} \times \hat{\mathbf{k}}_t \right) \otimes \left( \hat{\mathbf{z}} \times \hat{\mathbf{k}}_t \right), \tag{S1}$$

where $\mathbf{u} \otimes \mathbf{v}$ represents the tensor product of two generic vectors $\mathbf{u}$ and $\mathbf{v}$, $\hat{\mathbf{k}}_t = \mathbf{k}_t / |\mathbf{k}_t|$ (with $\mathbf{k}_t = k_x \hat{\mathbf{x}} + k_y \hat{\mathbf{y}}$ being the transverse wave vector), and $\sigma_L$ and $\sigma_T$ are the surface conductivities of the graphene sheet for longitudinal (with in-plane electric field parallel

---

to the wave vector **k**) and transverse (with electric field perpendicular to the wave vector **k**) excitations, respectively.

Based on the theoretical model reported in the Appendix C of Ref. [1], the collisionless nonlocal longitudinal intraband conductivity of graphene is given by:

$$\sigma_{\mathrm{L}}(\omega,k) = \frac{i\omega e^2}{\hbar^2} \frac{2\mu_{\mathrm{c}}}{\pi} \frac{1}{\omega\sqrt{\omega - v_{\mathrm{F}}k}\sqrt{\omega + v_{\mathrm{F}}k} + \omega^2 - v_{\mathrm{F}}^2 k^2}, \qquad (S2)$$

and the collisionless nonlocal transverse conductivity is given by

$$\begin{aligned}\sigma_{\mathrm{T}}(\omega,k) &= \frac{i\omega e^2}{\hbar^2} \frac{2\mu_{\mathrm{c}}}{\pi} \frac{\sqrt{\omega - v_{\mathrm{F}}k}\sqrt{\omega + v_{\mathrm{F}}k}}{\omega\left(\omega\sqrt{\omega - v_{\mathrm{F}}k}\sqrt{\omega + v_{\mathrm{F}}k} + \omega^2 - v_{\mathrm{F}}^2 k^2\right)} \\ &= \frac{\sqrt{\omega - v_{\mathrm{F}}k}\sqrt{\omega + v_{\mathrm{F}}k}}{\omega} \sigma_{\mathrm{L}}(\omega,k)\end{aligned}, \qquad (S3)$$

where $\omega$ is the oscillation frequency, $e$ is the absolute value of the electron charge, $\hbar$ is the reduced Planck constant ($\hbar = h/(2\pi)$), $\mu_{\mathrm{c}}$ is the chemical potential, $v_{\mathrm{F}} = c/300$ is the Fermi velocity, and $k = \sqrt{k_x^2 + k_y^2}$.

The effect of collisions can be taken into account with the phenomenological correction [1]

$$\sigma_l^{\mathrm{col}}(\omega,k,\tau) = -i\omega \frac{(1 + i/(\omega\tau))\chi_l(\omega + i\tau, k)}{1 + i/(\omega\tau)\chi_l(\omega + i\tau, k)/\chi_l(0,k)}, \quad l = \mathrm{L, T}, \qquad (S4)$$

where $\chi_l(\omega,k) = \sigma_l(\omega,k)/(-i\omega)$ is the susceptibility and $\tau$ is the relaxation time.

## *B. Reflection and transmission matrices*

Here we obtain general formulas for the reflection and transmission matrices ($\bar{\mathbf{R}}$ and $\bar{\mathbf{T}}$) of a generic 2D material with an electromagnetic response characterized by a surface conductivity tensor (Fig. S1) in terms of admittance matrices.

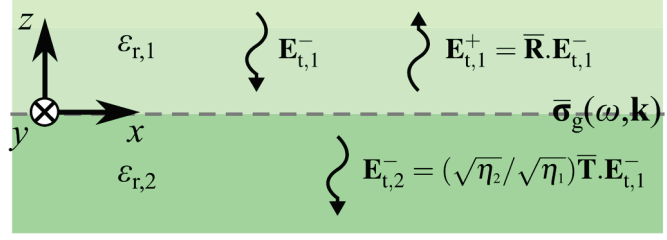

**Fig. S1.** Sketch of a graphene sheet surrounded by two dielectrics with relative permittivities $\varepsilon_{r,1}$ and $\varepsilon_{r,2}$ and wave impedances $\eta_1 = \eta_0/\sqrt{\varepsilon_{r,1}}$ and $\eta_2 = \eta_0/\sqrt{\varepsilon_{r,2}}$.

Following the approach of Refs. [2-3], we define the transverse fields as:

$$\mathbf{E}_t = \begin{pmatrix} E_x \\ E_y \end{pmatrix} \quad \text{and} \quad \mathbf{J} \cdot \mathbf{H}_t = \begin{pmatrix} 0 & 1 \\ -1 & 0 \end{pmatrix} \begin{pmatrix} H_x \\ H_y \end{pmatrix} = \begin{pmatrix} H_y \\ -H_x \end{pmatrix}. \tag{S5}$$

We introduce two admittance matrices $\overline{\mathbf{Y}}^{\pm}$ such that for plane waves propagating along the $+z$ and $-z$ directions one has:

$$\mathbf{J} \cdot \mathbf{H}_t^+ = \overline{\mathbf{Y}}^+ \cdot \mathbf{E}_t^+, \qquad \mathbf{J} \cdot \mathbf{H}_t^- = -\overline{\mathbf{Y}}^- \cdot \mathbf{E}_t^-. \tag{S6}$$

The matrices $\overline{\mathbf{Y}}^{\pm}$ depend on the considered material, on the frequency $\omega$, and on the transverse wave vector $\mathbf{k}_t$.

Let us consider the system illustrated in Fig. S1. By matching the tangential component of the electric field ($\mathbf{E}_{t,1}\big|_{z=0^+} - \mathbf{E}_{t,2}\big|_{z=0^-} = 0$) and by imposing the impedance boundary condition ($-\mathbf{J} \cdot (\mathbf{H}_{t,1}\big|_{z=0^+} - \mathbf{H}_{t,2}\big|_{z=0^-}) = \overline{\boldsymbol{\sigma}}_g \cdot \mathbf{E}_t$) at the interface [1, 4], it is found that:

$$(\overline{\mathbf{1}}_t + \overline{\mathbf{R}}) \cdot \mathbf{E}_{1,t}^- = \mathbf{E}_{2,t}^-, \qquad \overline{\mathbf{Y}}_1^- \cdot \mathbf{E}_{t,1}^- - \overline{\mathbf{Y}}_1^+ \cdot \mathbf{E}_{t,1}^+ - \overline{\mathbf{Y}}_2^- \cdot \mathbf{E}_{t,2}^- = \overline{\boldsymbol{\sigma}}_g \cdot \mathbf{E}_{t,2}^-. \tag{S7}$$

where $\overline{\mathbf{1}}_t = \hat{\mathbf{x}} \otimes \hat{\mathbf{x}} + \hat{\mathbf{y}} \otimes \hat{\mathbf{y}}$ is the transverse identity matrix.

Since for isotropic media $\overline{\mathbf{Y}}_i = \overline{\mathbf{Y}}_i^{\pm}$ ($i = 1, 2$), one obtains from (S7) that:

$$\overline{\mathbf{R}} = (\overline{\mathbf{Y}}_1 + \overline{\mathbf{Y}}_2 + \overline{\boldsymbol{\sigma}}_g)^{-1} \cdot (\overline{\mathbf{Y}}_1 - \overline{\mathbf{Y}}_2 - \overline{\boldsymbol{\sigma}}_g), \quad \text{and} \quad \overline{\mathbf{T}} = 2\sqrt{\frac{\eta_1}{\eta_2}} (\overline{\mathbf{Y}}_1 + \overline{\mathbf{Y}}_2 + \overline{\boldsymbol{\sigma}}_g)^{-1} \cdot \overline{\mathbf{Y}}_1 \tag{S8}$$

The admittance matrices $\overline{\mathbf{Y}}_i$ ($i = 1, 2$) are given by [2-3]:

$$\mathbf{Y}_i = \frac{1}{\eta_0 k_0 k_{z,i}^+}\begin{bmatrix} k_0^2\varepsilon_{r,i} - k_y^2 & k_x k_y \\ k_x k_y & k_0^2\varepsilon_{r,i} - k_x^2 \end{bmatrix}, \quad (S9)$$

where $k_{z,i}^+ = \sqrt{k_0^2\varepsilon_{r,i} - k_x^2 - k_y^2}$, $k_0 = \omega\sqrt{\varepsilon_0\mu_0}$ is the free-space wave number, and $\eta_0 = \sqrt{\mu_0/\varepsilon_0} = 120\pi\ [\Omega]$ is the free-space impedance.

The same result can be derived in a different way, as follows. Specifically, the reflection matrix (in the absence of a drift current) can be written as $\mathbf{R} = R_{TM}\left(\hat{\mathbf{k}}_t \otimes \hat{\mathbf{k}}_t\right) + R_{TE}\left(\hat{\mathbf{z}}\times\hat{\mathbf{k}}_t\right)\otimes\left(\hat{\mathbf{z}}\times\hat{\mathbf{k}}_t\right)$, with $R_{TM}$ and $R_{TE}$ the scalar electric reflection coefficients of the graphene sheet for plane waves incident with transverse magnetic (TM) and electric (TE) polarization. This decomposition is valid provided the surface conductivity has the structure $\bar{\boldsymbol{\sigma}}_g = \sigma_L\left(\hat{\mathbf{k}}_t \otimes \hat{\mathbf{k}}_t\right) + \sigma_T\left(\hat{\mathbf{z}}\times\hat{\mathbf{k}}_t\right)\otimes\left(\hat{\mathbf{z}}\times\hat{\mathbf{k}}_t\right)$

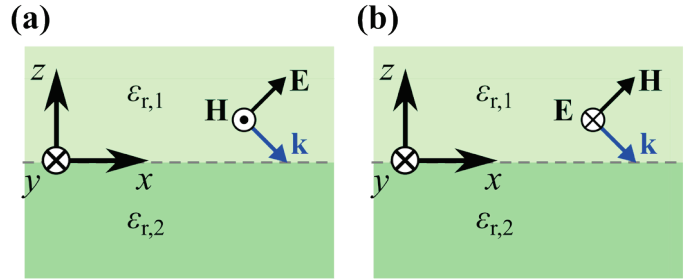

**Fig. S2.** Sketch of a graphene sheet illuminated by (a) a TM polarized wave and (b) a TE polarized wave.

Thus, $R_{TM}$ and $R_{TE}$ are coincident with the reflection coefficients for TM and TE polarized waves, for a plane wave illuminating an equivalent isotropic 2D material with surface conductivity $\sigma_L$ and $\sigma_T$, respectively. This observation implies that:

$$R_{TE}(\omega, k_x, k_y) = \frac{\gamma_1 - \gamma_2 + ik_0\eta_0\sigma_T}{\gamma_1 + \gamma_2 - ik_0\eta_0\sigma_T}$$

$$R_{TM}(\omega, k_x, k_y) = \frac{\gamma_2\varepsilon_{r,1} - \gamma_1\varepsilon_{r,2} + \gamma_1\gamma_2\dfrac{\sigma_L}{i\omega\varepsilon_0}}{\gamma_2\varepsilon_{r,1} + \gamma_1\varepsilon_{r,2} - \gamma_1\gamma_2\dfrac{\sigma_L}{i\omega\varepsilon_0}}, \quad (S10)$$

$\gamma_1 = \sqrt{k_x^2 + k_y^2 - \varepsilon_{r,1}(\omega/c)^2}$ and $\gamma_2 = \sqrt{k_x^2 + k_y^2 - \varepsilon_{r,2}(\omega/c)^2}$ are the propagation constants along $z$ in regions 1 and 2, respectively. It can be checked that the above formulas lead to an **R** that agrees exactly with Eq. S8.

## C. Fields radiated by an electric dipole

Here, we present the formulas for the electric field radiated by a Hertz electric dipole nearby a 2D material (e.g., a graphene sheet) [see Fig. S3]. The radiated and scattered fields are written in terms of Sommerfeld-type integrals.

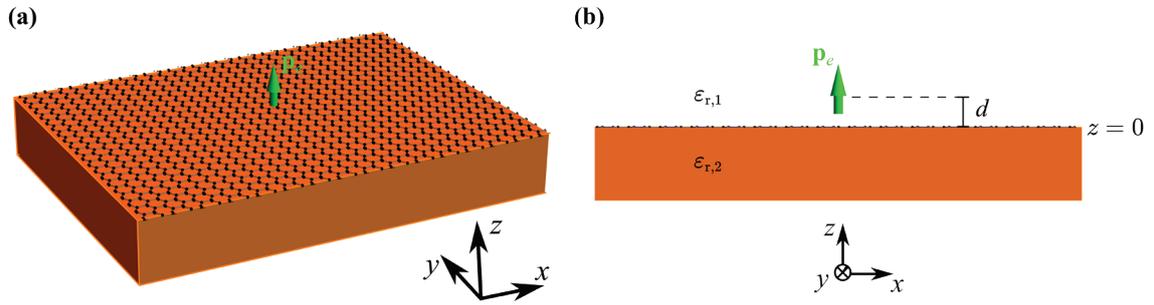

**Fig. S3.** An electric dipole stands above a graphene sheet. (a) Perspective view; (b) Front view.

The electromagnetic field in the region $z > 0$ where is located the dipole is given by the superposition of the primary field (i.e., the field radiated by the dipole) and the scattered field (i.e., the field scattered by the graphene sheet). The primary or radiated electric field is given by

$$\mathbf{E}^{inc} = \left(\nabla\nabla + k_0^2 \varepsilon_{r,1} \overline{\mathbf{1}}\right) \Phi_0 \cdot \frac{\mathbf{p}_e}{\varepsilon_0 \varepsilon_{r,1}}$$
$$\Phi_0 = \frac{e^{ik_0\sqrt{\varepsilon_{r,1}}|\mathbf{r}-\mathbf{r}'|}}{4\pi|\mathbf{r}-\mathbf{r}'|} = \frac{1}{(2\pi)^2} \iint dk_x dk_y \frac{e^{-\gamma_1|z-d|}}{2\gamma_1} e^{i(k_x x + k_y y)},$$
(S11)

Following the formalism of Ref. [5] and of the Appendix B of Ref. [6], the scattered electric field above the graphene sheet (i.e., the electric field reflected by the graphene sheet) can be written as:

$$\mathbf{E}^{ref} = \frac{1}{(2\pi)^2} \iint d^2\mathbf{k}_t \frac{e^{-\gamma_1(z+d)}}{2\gamma_1} e^{i\mathbf{k}_t \cdot \mathbf{r}} \mathbf{C}_1(\omega, \mathbf{k}_t) \cdot \frac{\mathbf{p}_e}{\varepsilon_0 \varepsilon_{r,1}},$$
(S12)

where

$$\mathbf{C}_1(\omega, \mathbf{k}_t) = \left[\overline{\mathbf{1}}_t + \hat{\mathbf{z}} \frac{i\mathbf{k}_t}{\gamma_1}\right] \cdot \overline{\mathbf{R}}(\omega, \mathbf{k}_t) \cdot \left[i\gamma_1 \mathbf{k}_t \hat{\mathbf{z}} + k_0^2 \varepsilon_{r,1} \overline{\mathbf{1}}_t - \mathbf{k}_t \mathbf{k}_t\right], \quad \text{(S13)}$$

$\overline{\mathbf{1}}_t = \hat{\mathbf{x}}\hat{\mathbf{x}} + \hat{\mathbf{y}}\hat{\mathbf{y}}$ is the transverse identity matrix, $\mathbf{k}_t = k_x \hat{\mathbf{x}} + k_y \hat{\mathbf{y}}$ is the transverse wave vector, $\gamma_1 = -ik_{z,1} = \sqrt{k_x^2 + k_y^2 - \omega^2 \varepsilon_0 \varepsilon_{r,1} \mu_0}$ is the propagation constant along z in region 1 ($z > 0$), and $\overline{\mathbf{R}}$ is the reflection matrix that relates the tangential (to the interface) components $x$ and $y$ of the reflected electric field to the corresponding $x$ and $y$ components of the incident field, $\begin{pmatrix} E_x^{\text{ref}} \\ E_y^{\text{ref}} \end{pmatrix} = \overline{\mathbf{R}}(\omega, \mathbf{k}_t) \cdot \begin{pmatrix} E_x^{\text{inc}} \\ E_y^{\text{inc}} \end{pmatrix}$, for the case of plane wave incidence.

On the other hand, the scattered electric field below the graphene sheet (i.e., the electric field transmitted through the graphene sheet) can be written as:

$$\mathbf{E}^{\text{tr}} = \frac{1}{(2\pi)^2} \iint \frac{1}{2\gamma_1} e^{-\gamma_1 d} e^{\gamma_2 z} e^{i\mathbf{k}_t \cdot \mathbf{r}} \mathbf{C}_2(\omega, \mathbf{k}_t) \cdot \frac{\mathbf{p}_e}{\varepsilon_0 \varepsilon_{r,1}} d^2 \mathbf{k}_t, \quad \text{(S14)}$$

where

$$\mathbf{C}_2(\omega, \mathbf{k}_t) = \left[\overline{\mathbf{1}}_t - \hat{\mathbf{z}} \frac{i\mathbf{k}_t}{\gamma_2}\right] \cdot \sqrt{\frac{\eta_2}{\eta_1}} \overline{\mathbf{T}}(\omega, \mathbf{k}_t) \cdot \left[i\gamma_1 \mathbf{k}_t \hat{\mathbf{z}} + k_0^2 \varepsilon_{r,1} \overline{\mathbf{1}}_t - \mathbf{k}_t \mathbf{k}_t\right], \quad \text{(S15)}$$

and $\gamma_2 = -ik_{z,2} = \sqrt{k_x^2 + k_y^2 - \omega^2 \varepsilon_0 \varepsilon_{r,2} \mu_0}$ is the propagation constant along z in region 2 ($z < 0$), and $\overline{\mathbf{T}}$ is the transmission matrix that relates the tangential (to the interface) components $x$ and $y$ of the transmitted electric field to the corresponding $x$ and $y$ components of the incident field, $\begin{pmatrix} E_x^{\text{tr}} \\ E_y^{\text{tr}} \end{pmatrix} = \sqrt{\frac{\eta_2}{\eta_1}} \overline{\mathbf{T}}(\omega, \mathbf{k}_t) \cdot \begin{pmatrix} E_x^{\text{inc}} \\ E_y^{\text{inc}} \end{pmatrix}$, for the case of plane wave incidence.

In the Galilean model, the reflection and transmission matrices in the presence of a drift-current bias along the *x*-direction are related to the no-drift matrices derived in Sect. A as $\tilde{\mathbf{R}}(\omega,k_x,k_y) = \bar{\mathbf{R}}(\tilde{\omega},k_x,k_y)$ and $\tilde{\mathbf{T}}(\omega,k_x,k_y) = \bar{\mathbf{T}}(\tilde{\omega},k_x,k_y)$, where $\tilde{\omega} = \omega - k_x v_0$ is the Doppler-shifted frequency and $v_0$ is the drift velocity of the electrons on the graphene sheet along the *x*-direction.

## D. Detailed study of the SPPs excited by an electric dipole emitter

Figure S4 present time snapshots of the SPPs propagating on a drift-current biased graphene deposited on a SiC substrate (analogous to Fig. 5(a) of the main text) and excited by short electric dipoles with different dipole moments $\mathbf{p}_e$.

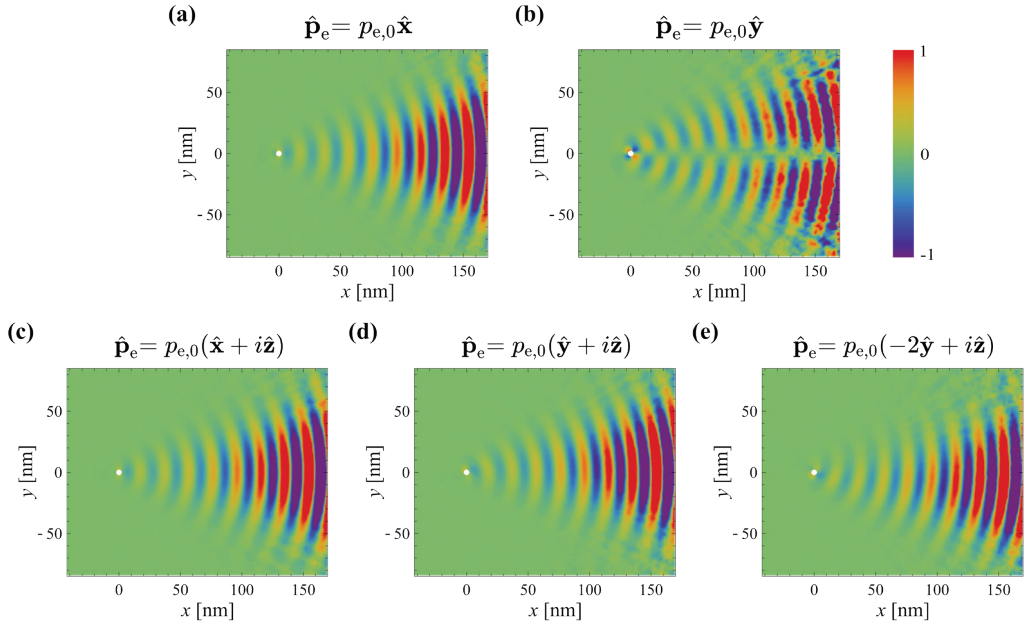

**Fig. S4.** (a-e) Time snapshots of the *x*-component of the electric field $E_x$ (in arbitrary unities) on the surface of the graphene sheet for electric dipoles placed 1 nm above the interface, calculated using the Galilean model with $v_0 = 0.6 v_F$. The electric dipole moment of each dipole is indicated on the top of each panel. In all the panels the frequency of operation is $\omega/(2\pi) = 25$ THz, $\mu_c = 0.1$ eV, $\tau = 1.7$ ps, $\varepsilon_{r,1} = 1$, and $\varepsilon_{r,2} = \varepsilon_{r,\text{Sic}}$.

Figure S4 shows that the graphene-SiC system enables SPP amplification for linearly polarized horizontal dipoles, as well as for dipoles with circular and elliptical polarizations, similar to that shown in Fig. 5(a) of the main text for a linearly polarized vertical dipole.